\documentclass[letters]{aa} 
\usepackage{newtxtext,newtxmath}
\usepackage{xspace,graphicx}

\newcommand{\spt}{SPT\,0346-52\xspace}

\newcommand\textlcsc[1]{\textsc{\MakeLowercase{#1}}}
\newcommand{\cii}{[C\,II]\xspace}

\begin{document}

\title{Detection of a high-redshift molecular outflow in a primeval hyperstarburst galaxy}

\author{
G.~C. Jones\inst{1,2}\thanks{E-mail: gj283@cam.ac.uk}
\and R. Maiolino\inst{1,2} 
\and P. Caselli\inst{3}
\and S. Carniani\inst{4}
}

\institute{
Cavendish Laboratory, University of Cambridge, 19 J. J. Thomson Ave., Cambridge CB3 0HE, UK
\and
Kavli Institute for Cosmology, University of Cambridge, Madingley Road, Cambridge CB3 0HA, UK
\and
Centre for Astrochemical Studies, Max-Planck-Institute for Extraterrestrial Physics, Giessenbachstrasse 1, 85748 Garching, Germany
\and
Scuola Normale Superiore, Piazza dei Cavalieri 7, I-56126 Pisa, Italy
}

\date{Received XXX; accepted XXX}

\abstract{
We report the discovery of a high-redshift, massive molecular outflow in the starburst galaxy SPT0346-52 ($z=5.656$) via the detected absorption of high-excitation water transitions (H$_2$O $4_{2,3}-4_{1,4}$ and H$_2$O $3_{3,0}-3_{2,1}$) with the Atacama Large Millimeter/submillimeter Array (ALMA). The host galaxy is one of the most powerful starburst galaxies at high redshift (star formation rate; SFR $\sim3,600$\,M$_\odot\,$year$^{-1}$), with an extremely compact ($\sim320$\,pc) star formation region and a star formation rate surface density ($\Sigma_{SFR}\sim5,500$\,M$_{\odot}~$year$^{-1}~$kpc$^{-2}$)  five times higher than `maximum' (i.e. Eddington-limited) starbursts, implying a highly transient phase. The estimated outflow rate is  $\sim500$\,M$_{\odot}$year$^{-1}$, which is much lower than the SFR, implying that in this extreme starburst the outflow capabilities saturate and the outflow is no longer capable of regulating star formation, resulting in a runaway process in which star formation will use up all available gas in less than 30\,Myr. Finally, while previous kinematic investigations of this source revealed possible evidence for an ongoing major merger, the coincidence of the hyper-compact starburst and high-excitation water absorption indicates that this is a single starburst galaxy surrounded by a disc.
}

\keywords{Galaxies: high-redshift - Galaxies: starburst - ISM: jets and outflows}

\maketitle

\section{Introduction} 
Massive galactic outflows in the early universe have been invoked as a key mechanism to regulate or even quench star formation in galaxies \citep[e.g.][]{mura15,hayw17}, and are therefore possibly responsible for the population of massive, passive, and old galaxies, some of which are already in place at $z\sim2-4$ \citep[e.g.][]{glazebrook17,schreiber18,morishita19,merlin19,santini19}. 
Massive, quasar-driven outflows have been detected at $z\sim6$, but they have turned out not to be as effective as expected (\citealt{maio12,cico15,bisc19}).
Massive outflows driven by star formation (either by supernovae or radiation pressure) have been more difficult to detect at high $z$. While the ionised component has been detected in large samples of high-$z$ star forming galaxies, the cold molecular or atomic component (which generally accounts for most of the mass in the outflows) has been much more difficult to trace and there are only a few detections at high $z$ reported to date (\citealt{geor14,spil18}).

Cold outflows have been detected in local galaxies using a number of tracers, including CO (e.g. \citealt{cico14,flue18}), \cii (e.g. \citealt{maio12,cico15,janssen16,bisc19}), OH (e.g. \citealt{fisc10,stur11,gonz17}), and H$_2$O (e.g. \citealt{gonz10}). Within this context, high-level water transitions are particularly interesting, as they trace gas that must be very dense and warm in order to populate these levels
(e.g. \citealt{apos19}), which are typically found in the cores of compact starbursts. 

In order to explore the dense and warm molecular outflow in the starburst galaxy SPT 0346-52 at $z=5.656$, we used band 7 of the Atacama Large Millimeter/submillimeter Array (ALMA) to observe the two high-level water transitions H$_2$O $4_{2,3}-4_{1,4}$ ($\nu_{\rm rest}=$2264.14965\,GHz, E$_{\rm U}\sim432$\,K) and H$_2$O $3_{3,0}-3_{2,1}(\nu_{\rm rest}=2196.345756$\,GHz, E$_{\rm U}\sim410$\,K). 
SPT-S J034640-5204.9 (hereafter \spt) is a strongly lensed dusty starforming galaxy (DSFG) first studied in the ALMA survey of \citet{weis13} and \citet{viei13} who targeted bright sources detected by the South Pole Telescope survey. Detailed lens modelling shows that the galaxy is magnified by $\mu=5.6\pm0.1$ \citep{spil16}, while a source-plane reconstruction reveals indications for either an ongoing major merger or a disturbed rotating disc \citep{spil15}. Spectral energy distribution (SED) modelling yields a massive star formation rate (SFR$=3600\pm300$\,$M_{\odot}$\,year$^{-1}$; \citealt{ma16}) and further observations have found substantial \cii ($\rm L_{[C\,II]}=(5.0\pm0.7)\times10^{10}\,L_{\odot}$; \citealt{gull15}) and CO ($\rm L_{CO(2-1)}=(2.4\pm0.2)\times10^{8}\,L_{\odot}$; \citealt{arav16}) emission. However,  X-ray, radio, and H$_2$O observations indicate that this source is mainly powered by star formation, with no evidence for a powerful active galactic nucleus (AGN) at any wavelength (\citealt{ma16,apos19}).

In this work, we present new ALMA observations of \spt in band 7, further constraining the size of the compact starburst and revealing H$_2$O $4_{2,3}-4_{1,4}$ and $3_{3,0}-3_{2,1}$ absorption, which are interpreted as signatures of massive outflows. We assume ($\Omega_{\Lambda}$,$\Omega_{m}$,h)=(0.692, 0.308, 0.678) throughout \citep{plan16}. At this distance, 1\,arcsecond corresponds to 6.033 proper kpc at the redshift of SPT 0346-52 ($z=5.656$).

\section{Observations and data reduction}
Our ALMA band 7 observations were taken between 4 and 11 October, 2016, using 40-43 antennas in configuration C40-7.
Out of 3.23\,hours of total observation time, 1.95\,hours were on-source. 
The complex gain, bandpass, and flux calibrators were J0425-5331, J0538-4405, and J0334-4008, respectively. 
These data were calibrated by ALMA staff following the standard pipeline. 

Our frequency range was covered by two sidebands, each composed of two spectral windows (SPWs) made of 128 channels, each of them 15.625\,MHz wide.
Both sidebands are tuned to the redshifted frequencies of water lines, specifically: H$_2$O 4$_{2,3}-4_{1,4}$  (redshifted to $340.167$\,GHz) and H$_2$O $3_{3,0}-3_{2,1}$ (redshifted to $329.980$\,GHz).

Continuum subtraction was performed in \textit{uv}-space (CASA uvcontsub) using all line-free channels (i.e. excluding $\pm750$\,km\,s$^{-1}$ from the expected frequency, assuming $z=5.656$; \citealt{arav16}).
While the continuum in each sideband changes by $\sim1-2\%$ over its corresponding frequency range, we still assume a flat continuum. Continuum subtraction that allowed for a slope in the continuum resulted in nonphysical slopes, and therefore undersubtraction of continuum. 
We emphasise that the water absorption lines in both the northern and southeastern components are detected clearly regardless of the continuum subtraction method.

Using the CASA task tclean, the continuum-subtracted data in each sideband were used to create spectral data cubes with channels of 15.6\,MHz (14\,km\,s$^{-1}$) and natural weighting in order to balance the RMS noise level and resolution.
Since the atmospheric transmission varies strongly across the lower band (i.e. between $0.3$ and $0.6$ for $2.0$\,mm precipitable water vapour), the RMS noise level per channel ranged from $0.2$ to $0.5$\,mJy\,beam$^{-1}$. 
The RMS noise level of the upper band is relatively constant at $0.2$\,mJy\,beam$^{-1}$ per channel.
A continuum image was created using the line-free channels that were used for continuum estimation, multifrequency synthesis, and natural weighting. In order to maximise signal recovery, the image was cleaned interactively, resulting in a restoring beam of $0.14''\times0.12''$ at a position angle of $\sim-50^{\circ}$, and an RMS noise level of $0.25$\,mJy\,beam$^{-1}$.

\section{Results and evidence for molecular outflow}

\begin{figure}
\centering
\includegraphics[width=0.95\columnwidth]{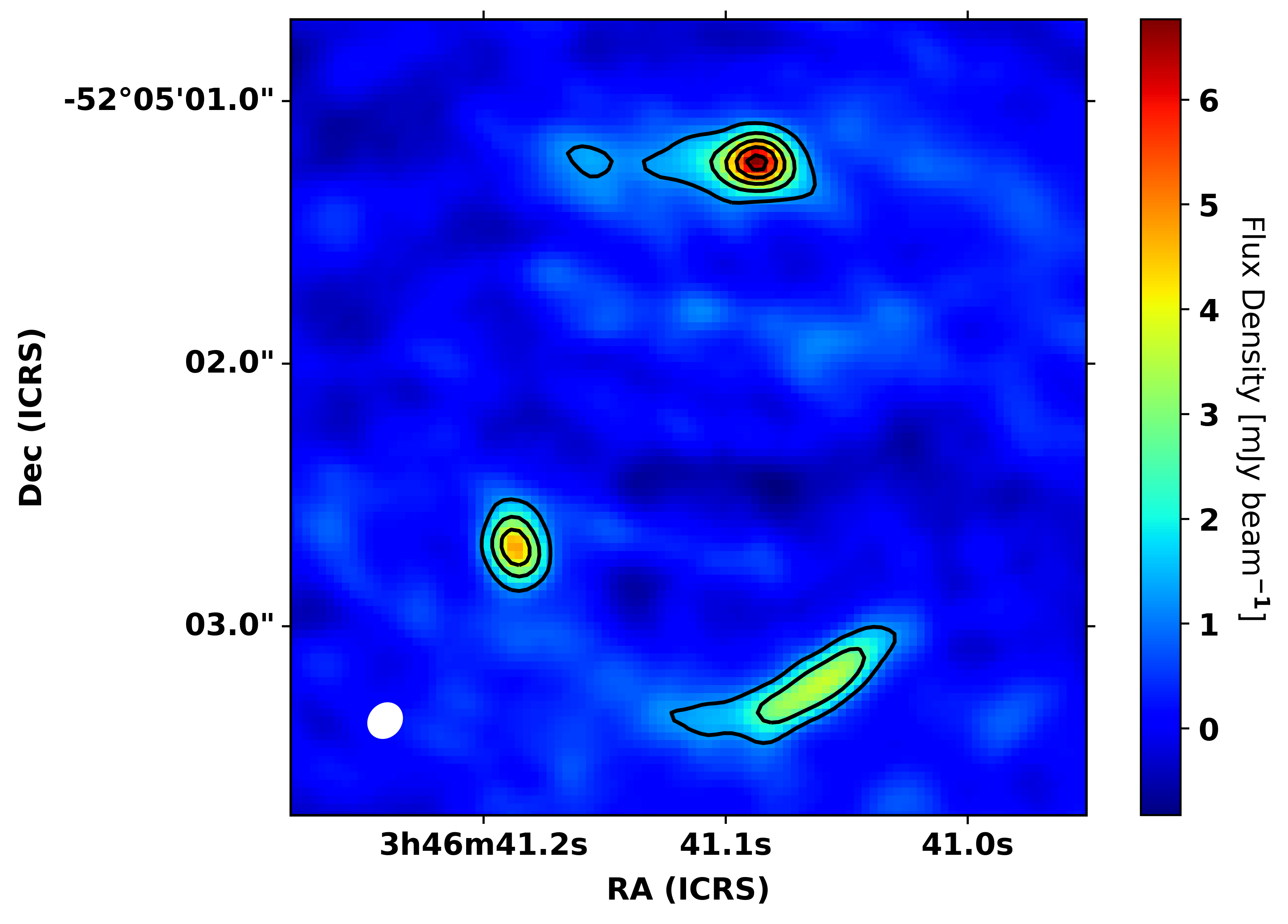}
\includegraphics[width=0.8\columnwidth]{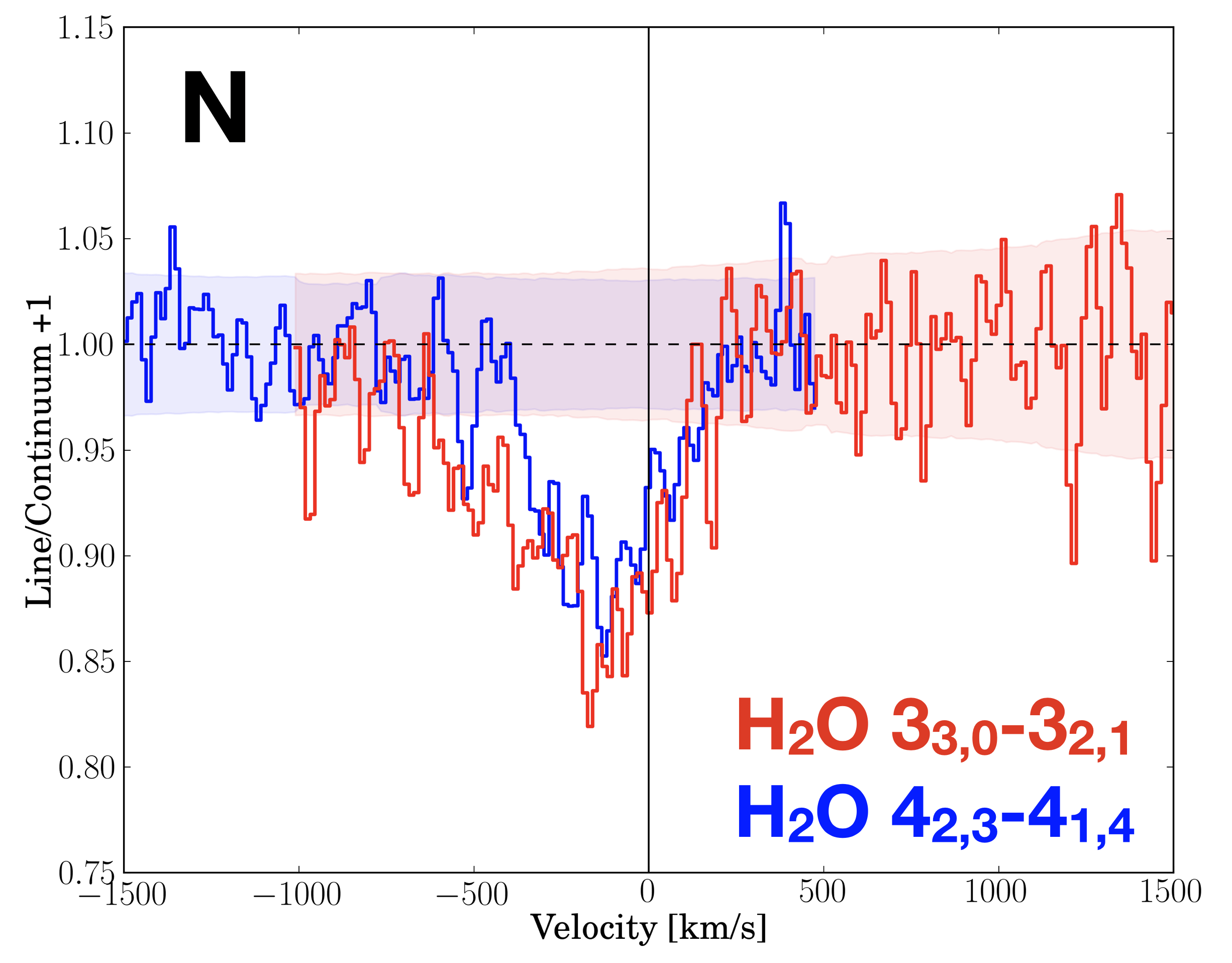}
\includegraphics[width=0.8\columnwidth]{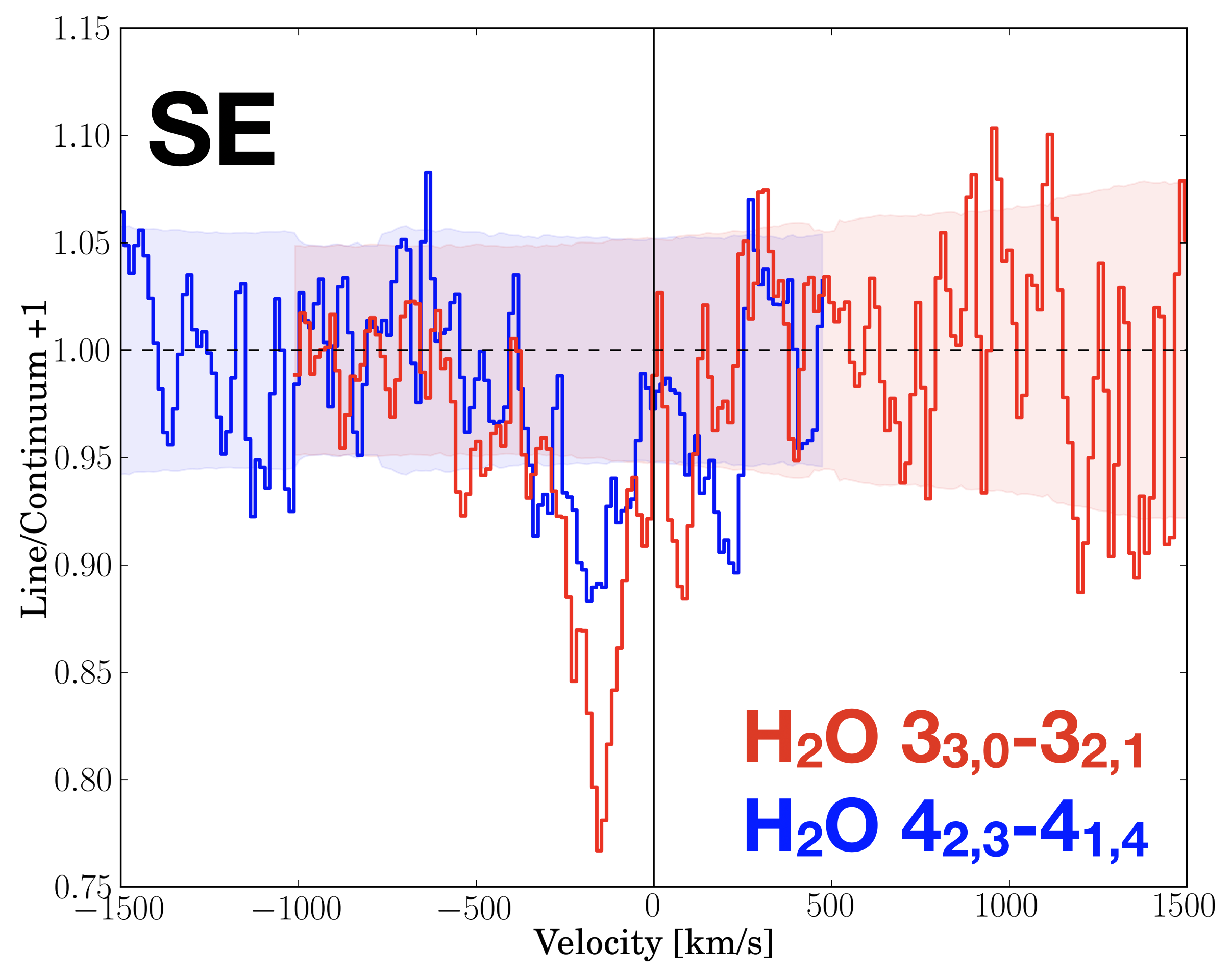}
%
\caption{ALMA Band 7 continuum emission and H$_2$O absorption. 
Top: ALMA band-7 continuum emission. Contours are in steps of $5\times(\sigma=0.25$\,mJy\,beam$^{-1})$. 
Centre: H$_2$O 3$_{3,0}$-3$_{2,1}$ and H$_2$O 4$_{2,3}$-4$_{1,4}$ absorption profiles for the northern (strongest) lensed image. The CO-derived redshift of the host galaxy is denoted by a vertical line.
Bottom: H$_2$O 3$_{3,0}$-3$_{2,1}$ and H$_2$O 4$_{2,3}$-4$_{1,4}$ absorption profiles for the southeastern lensed image. The RMS noise level spectra are shown as shaded areas.}
\label{contspec}
\end{figure}

The underlying continuum emission (at $\rm \lambda _{rest}\sim 130\mu m$, due to emission from warm dust heated by the radiation of young stars) is clearly detected and resolved in three lensed images (top panel of Figure \ref{contspec}), consistent with previous observations (\citealt{viei13,gull15,spil15}).
We detect clear absorption associated with the two water transitions at the location of the two strongest continuum images of the galaxy (i.e. the northern and southeastern; middle and bottom panels of Figure \ref{contspec}), but no absorption feature is detected in the third image, which however has much lower surface brightness in continuum and therefore the associated spectrum is more noisy. At the northern and southeastern component the water absorption is centred on the peak of the two continuum images and is not resolved.

\begin{figure}
\centering
\includegraphics[width=\columnwidth]{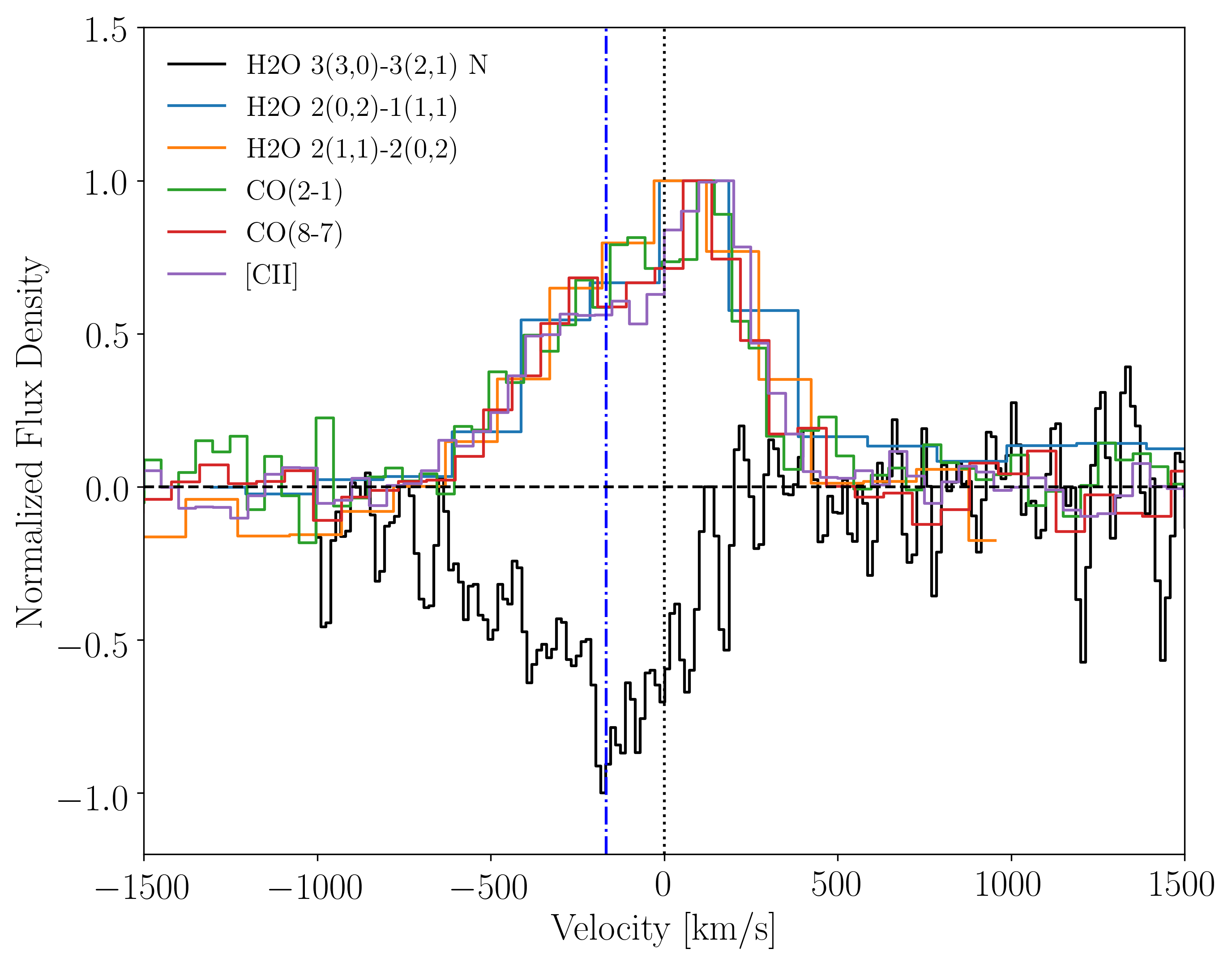}
\caption{Water absorption profile for the northern lensed image compared with the total  [CII]\,158$\mu$m \citep{litk19}, CO(8-7) \citep{dong19}, CO(2-1) \citep{arav16}, and water emission profiles integrated over all three images. The H$_2$O $2_{0,2}-1_{1,1}$ and $2_{1,1}-2_{0,2}$ emission profiles were taken from ALMA archival data (2013.1.00722.S and 2015.1.00117.S, respectively). All spectra were normalised to their peak/dip and the velocity zero scale is set to $z_{CO}=5.6559$, shown by the black dotted line. The minimum of the water trough ($z=5.6522$) is shown by the blue dash-dotted line.}
\label{overplot}
\end{figure}

To ease interpretation of the water absorption line features, we primarily consider the strongest absorption, H$_2$O $3_{3,0}-3_{2,1}$ in the northern image.
The bottom of the absorption feature is blueshifted by $\sim170$\,km\,s$^{-1}$ with respect to the local standard of rest traced by the centroid of the emission transitions of CO (\citealt{arav16,dong19}), [CII] (\citealt{litk19}), and water emission lines associated with lower-energy transitions, as illustrated in Figure \ref{overplot}. 
The data for each of these emission profiles were taken from the ALMA pipeline products, and the emission profiles shown here were extracted over an aperture containing all three images, while the water absorption profile was extracted from a compact region. 
The profile of the absorption is clearly asymmetric, with a blue wing extending to $\rm -500$\,km\,s$^{-1}$ with respect to the absorption dip.

It should be noted that after taking velocity-dependent differential lensing into account, the intrinsic CO(8-7) spectrum was shown to be more symmetric \citep{dong19}. However, this differential lensing effect can only affect the extended, low-excitation emission lines, but not the compact (unresolved), high-excitation H$_2$O absorption profile. 

An asymmetric blueshifted (by 500\,km\,s$^{-1}$) absorption profile as seen here, especially for a transition that requires extreme gas conditions (extremely high densities, high temperatures, and high column densities, which cannot be associated with tidal features or satellite galaxies), can only be explained in terms of a dense molecular outflow along the line of sight (\citealt{gonz10,gonz17,fisc10,stur11,spil18}). 

\section{Analysis}

\subsection{Hyperstarburst nature of \spt}

The far-infrared (FIR) continuum emission (tracing star formation) in the northern image is spatially resolved and has a beam-deconvolved size of $(0.34\pm0.03)''\times(0.19\pm0.02)''$. Taking the velocity-independent continuum magnification factor \citep{spil16} of $\mu=5.6\pm0.1$
into account, this corresponds to a source-plane half width at half max (HWHM) of $0.32\pm0.02$\,kpc. Using the SFR derived from a fit to SED constructed of multiple IR/submillimeter continuum measurements \citep{ma16} (SFR$=(3.6\pm0.3)\times10^3$\,M$_{\odot}$\,year$^{-1}$) and including a factor of 0.5 to scale the total SFR to the fraction of FIR flux contained within the full width at half max (FWHM)\footnote{This factor of 0.5 was derived by numerically integrating an elliptical Gaussian with unity amplitude over all space, and dividing this value by the integral of the same elliptical Gaussian over all space where its value was $\ge0.5$.}, we find a SFR density of $\Sigma_{SFR}=(5.5\pm0.9)\times10^3$\,M$_{\odot}$\,year$^{-1}$\,kpc$^{-2}$. This value is larger than the previously derived $\sim4.2\times10^3$\,M$_{\odot}$\,year$^{-1}$\,kpc$^{-2}$ \citep{heza13} and $(1.5\pm0.1)\times10^3$\,M$_{\odot}$\,year$^{-1}$\,kpc$^{-2}$ \citep{ma16}, due to our new constraints from our high-resolution imaging. This surface density of star formation is a factor of five times higher than that of `maximum' (i.e. Eddington-limited) starbursts ($\Sigma_{SFR,EL}\sim10^3$\,M$_{\odot}$\,year$^{-1}$\,kpc$^{-2}$; \citealt{thom05}).

This extreme surface density of SFR may potentially be an indication that the IR radiation is actually powered by an AGN, however there is no evidence for an AGN at any wavelength, not even in the hard X-ray Chandra observations, which at this redshift sample energies out to 60~keV \citep{ma16}. In principle, the AGN could be completely Compton thick (i.e. absorbed by a column of gas larger than $10^{25}~$cm$^{-2}$). However, the SED, including IR bands, is better fit using a pure starburst model than a composite starburst and AGN model \citep{ma16}. Moreover, the lack of radio emission  is fully consistent with what is expected by the SFR, and leaves little room for any contribution to the bolometric luminosity from a heavily Compton thick, obscured quasar.

\subsection{Kinematic nature of \spt}

Previous observations of the extended [CII], CO, and H$_2$O emission from SPT 0346-52 (\citealt{litk19,dong19,apos19}) and their associated source-plane reconstruction suggested evidence for an ongoing merger of two galaxies, separated by $\sim500$\,km\,s$^{-1}$ and 1\,kpc in the source plane. Namely, they revealed asymmetric double-peaked emission, a non-uniform velocity gradient, and a `bridge' of emission connecting the two sources. 

However, a full multi-channel gravitational lens analysis revealed that while the CO(8-7) profile appears asymmetric, this is only due to differential lensing, and it is intrinsically symmetric \citep{dong19}. In addition, a simple rotating disc is expected to present two emission peaks at high velocity, with a lower-significance bridge connecting them (e.g. \citealt{nort19}). With this in mind, the current kinematics of this source are not sufficient to distinguish between a double merger and a single rotator (\citealt{bois11,simo19}). 

\begin{figure}
\centering
\includegraphics[width=0.8\columnwidth]{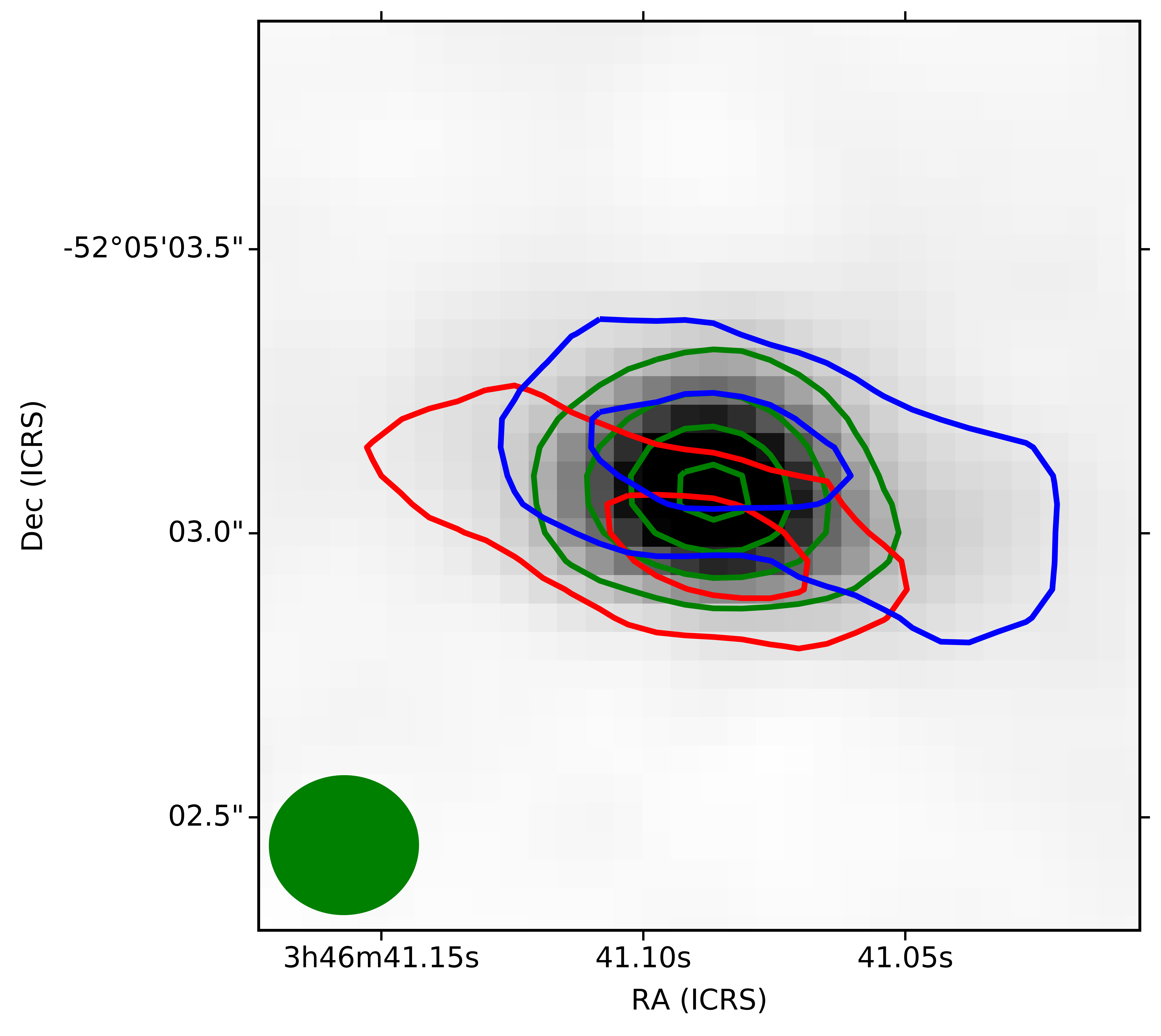}
\caption{Red and blue halves of the [CII] emission in red and blue contours respectively, together with the rest-frame $158\,\mu$m continuum in green and greyscale \citep{litk19}. Data were taken from the ALMA archive and passed through the automatic CASA calibration pipeline. A flat continuum was fit to the \textit{uv}-space data using all line-free channels, and this continuum was subtracted using the CASA task \textlcsc{uvcontsub}. The continuum image was created with natural weighting using the continuum model subtracted from the full cube, while the line cube was created with the continuum-free visibilities. No lensing correction has been applied (i.e. images are image-plane). [CII] contours are shown at $5,10\sigma$, where $1\sigma=$0.13\,Jy\,beam$^{-1}$\,km\,s$^{-1}$, and 0.16\,Jy\,beam$^{-1}$\,km\,s$^{-1}$ for red and blue, respectively. Continuum contours are shown at $10,20,30,40\sigma$, where $1\sigma=$0.47\,mJy\,beam$^{-1}$. The continuum  peak is clearly located between the red and blue [CII] peaks.
The synthesised beam of the continuum image is shown by the filled green ellipse in the lower left.}
\label{ciicont}
\end{figure}

One point that breaks this degeneracy is the finding that the very compact continuum emission tracing the hyper-Eddington starburst is located between the two [CII] peaks (Figure \ref{ciicont}), coincident with the peak of velocity dispersion. 
Moreover, our detection of high-excitation, compact water absorption is coincident with this continuum peak. 
The small spatial extent of absorption requires a compact continuum source.
The water absorption lines detailed in this work (H$_2$O $4_{2,3}-4_{1,4}$ and $3_{3,0}-3_{2,1}$)
are associated with ortho-H$_2$O upper levels that are at an energy of $\rm E_u/k_B\sim 400$\,K above ground, and therefore they require high temperatures, high gas densities ($\rm n_H>10^6\,cm^{-3}$), and high column densities ($\rm N_H>10^{24}\,cm^{-2}$) to be observable (\citealt{vand07,gonz12}). Furthermore, the lower levels, which must be populated in order to see our lines in absorption, have $\rm E_l/k_B\sim300$\,K, underlining the need for high-excitation gas to be able to see our lines.
These conditions (extremely compact source, hyper-Eddington starburst, and extremely high column densities of warm gas) are seen only in the cores of some extreme ultra-luminous infrared galaxies (ULIRGs) (\citealt{gonz10,gonz12}), and cannot be associated with a tidal bridge of diffuse gas between two merging galaxies \citep{schi16}.

These findings show that the merger interpretation cannot hold. Due to the high redshift and excitation of this source, it is likely a rotating disc undergoing a number of minor mergers, resulting in a galaxy similar to the simulated $z=6.18$ ``disturbed disk'' of \citet{koha19}. This would explain the strong, compact central continuum emission and central dense and warm outflow, as associated with the starbursting core of the galaxy. In addition, the asymmetric double horned profile (see figure 6 of Kohandel et al.), which however becomes symmetric when corrected for differential lensing \citep{dong19}, and PV diagrams of [CII] \citep{litk19} can be explained in terms of a galactic disc (or even a ring). 

\subsection{Outflow model}

In order to interpret the water absorption profile, we created a simple spherical outflow model and examined its predicted spectrum. In this model, a spherical continuum emitter is surrounded by a shell of outflowing gas which we assume to feature a constant mass outflow rate (i.e. related velocity and density profiles). The optical depth at each radius is then determined and used to derive the fraction of emitted photons at a given frequency blocked by each parcel of gas. By performing this calculation at all locations in the outflowing shell, we create an absorption spectrum normalised by the total continuum level. Assuming a range of shell widths and a velocity and density profile, a Bayesian inference code (MultiNest; \citealt{fero08}) was used to explore a wide parameter space (i.e. outflow velocity dispersion, inner optical depth, and launch velocity) to find the best-fit parameters\footnote{The spherical outflow model used to fit the observed spectrum is available at https://github.com/gcjones2/SphOut. See Appendix for more details.}.

Since the current data suggest that this source is a single, rotating galaxy rather than a pair of merging galaxies, we adopt $z_{CO}$ as the source redshift (where $z_{CO}$ is taken from the centroid of the CO profile). Exploring a range of shell widths (i.e. $dR/R_{CONT}=3$, 19, and 10, following the geometries obtained by \citet{gonz10,gonz12,gonz17}, and assuming that the CO-based redshift traces the rest-frame of the galaxy, these parameters yield possible total mass outflow rates in the range $ \dot{M}\sim100-900\,$M$_{\odot}\,$year$^{-1}$, with an average value of $ \dot{M}\sim500\,$M$_{\odot}\,$year$^{-1}$.

\begin{figure}
\centering
\includegraphics[width=\columnwidth,trim=0 0 0 0, clip]{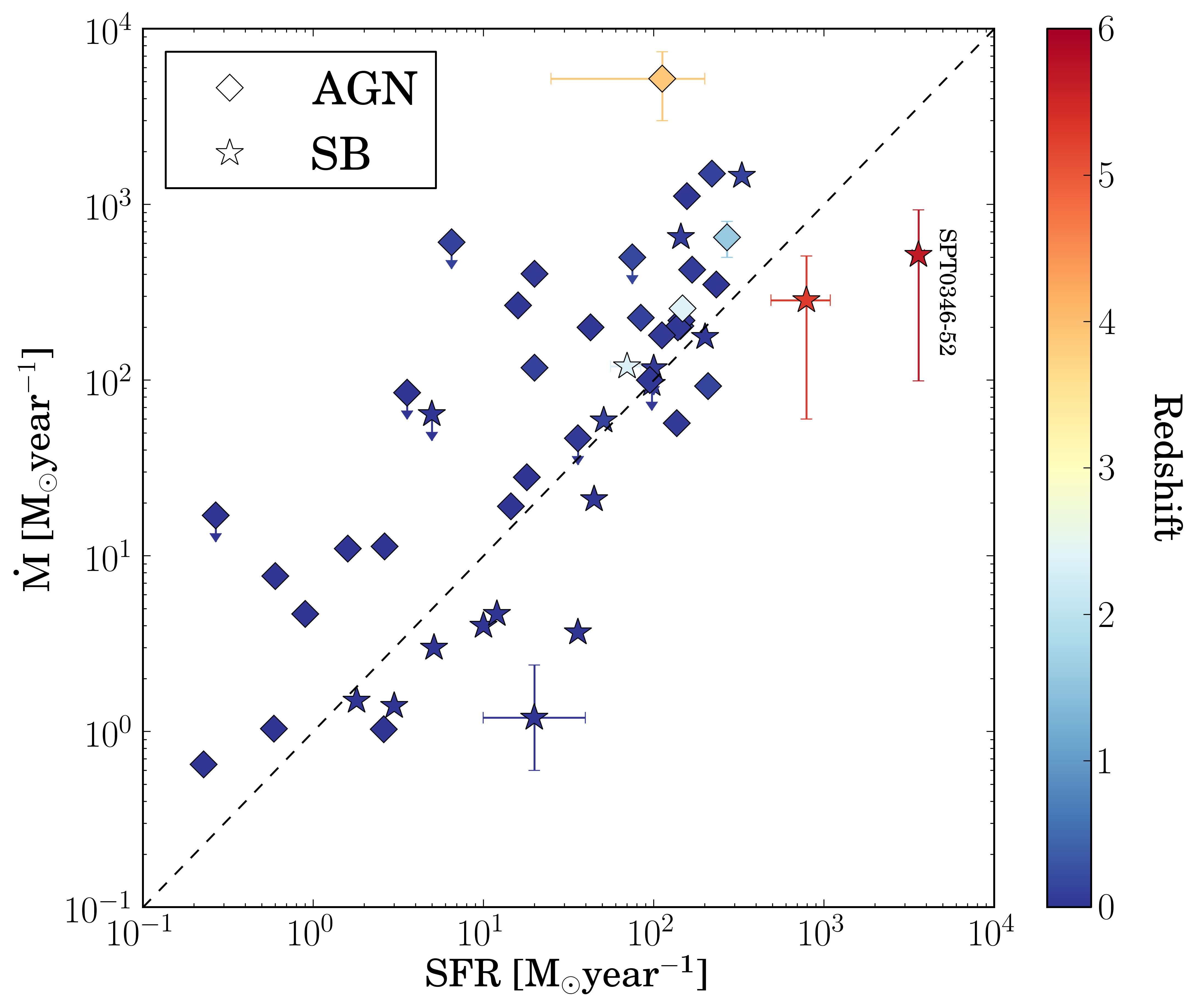}
\caption{Molecular outflow rate as a function of SFR for galaxies with detected molecular outflows.
The range and average of best-fit outflow rates of SPT 0346-52 for three possible shell widths are shown, assuming $z_{CO}$ and SFR$_{SED}=(3.6\pm0.3)\times10^3$\,M$_{\odot}$\,year$^{-1}$ \citep{ma15}.
The molecular outflow detected in this work is in the most powerful starburst. 
Additional data include both low-redshift sources \citep{flue18} and the higher-redshift galaxies XID2028 ($z=1.593$; \citealt{brus18}), SMM J2135 ($z=2.326$; \citealt{geor14}), zC400528 ($z=2.387$; \citealt{herr19}), APM08279+5255 ($z=3.912$; \citealt{feru17}), and SPT2319-55 ($z=5.293$; \citealt{spil18}). 
A representative uncertainty for all low-redshift sources ($\pm0.3$\,dex) is shown for one such source.
Outflows driven by AGNs are shown by diamonds, while those driven by starbursts are shown by star symbols. Each object is coloured according to its redshift.}
\label{sphout}
\end{figure}

When the results are compared to local galaxies and other high-redshift outflows, it is apparent that the molecular outflow in SPT0346-52 is the one observed in the galaxy with the highest SFR, by a large factor (Figure \ref{sphout}). 

Outflow rate can also be quantified in terms of the ratio of the rate of mass expelled by the outflow to the SFR (i.e. the outflow mass loading factor, $\eta$). 
If we take the $z_{CO}$ mass outflow rate values and assume SFR$_{IR}=(3.6\pm0.3)\times10^3$\,M$_{\odot}$\,year$^{-1}$ \citep{ma16}, then we find a mass loading factor of $\rm \eta\sim0.03-0.28$. 
An earlier SED fit \citep{ma15} yielded a larger value of SFR$_{SED}=(4.5\pm1.0)\times10^3$\,M$_{\odot}$\,year$^{-1}$, which would result in an even smaller mass loading factor of $\eta=\dot{M}$/SFR$=0.02-0.22$. For all assumed shell widths and SFRs, we find a significant outflow, with a mass-loading factor well below unity.

One may argue that, given that the CO, [CII], and (low-level) H$_2$O emission lines trace the rotating disc, their centroid may not provide an accurate measurement of the systemic velocity. One possibility is that the dip of the water absorption traces the systemic velocity of the core of the galaxy.
If one adopts the dip of the water absorption as the rest-frame of the galaxy, then the inferred outflow rate drops
by more than one order of magnitude, making an even stronger case that in this galaxy the mass-loading factor of the starburst-driven outflow is well below unity.

Outflows of local star forming galaxies show a mass-loading factor of order unity while those hosting a powerful AGN often have a loading factor significantly higher than unity (\citealt{cico14,flue18}), as illustrated in Fig.~\ref{sphout}.
The outflow loading factor in SPT0346-52 is not well constrained (i.e. $\eta=0.03-0.28$, possibly even lower),
but it is well below unity. 
Most likely, for this hyperstarburst galaxy, whose $\rm \Sigma_{SFR}$ is more than five times above the Eddington limit, its capability of driving an outflow has saturated and the outflow rate does not follow the same scaling with SFR as in other galaxies. The implication is that the outflow does not have the same `regulatory' effect as in other less extreme star forming galaxies (lower SFR, less compact, and lower $\Sigma _{SFR}$). As a consequence this galaxy is probably in a runaway phase, in which star formation is proceeding at a high rate and high efficiency, consuming all of the available gas in less than $\sim30$\,Myr \citep{litk19}. If there is no accretion of fresh gas from the intergalactic medium (IGM) then this galaxy will quickly turn into a passive galaxy, which will evolve into the population of passive galaxies observed to be already in place at $z\sim2-4$.

It is interesting that, contrary to many standard scenarios, `uneffective' outflows (uneffective in their `ejective' mode) could be a new route to rapidly quench star formation in a galaxy, in the sense that this type of outflows allows star formation to enter into a runaway process and to consume gas at an even higher rate than usual. The key role of the outflow may however be to dump energy into the halo, thereby keeping it hot, and preventing the accretion of fresh cold gas which could restart star formation.

\section{Conclusions}

Here we present new ALMA observations in band 7 of \spt, a strongly 
lensed starburst galaxy at $z=5.656$, characterised by an SFR of $3,600$~M$_{\odot}~$year$^{-1}$.

Thanks to our higher angular resolution
relative to previous observations, we can measure more accurately measure
the size of the continuum emission (and therefore of the star forming region), which turns out to have a half-light radius of $320\pm20$~pc. This results in an extremely high surface density of star formation rate of $\Sigma _{SFR}=5,500~$M$_{\odot}~$year$^{-1}~$kpc$^{-2}$, which is more than five times higher than in a `maximum' starburst (i.e. an Eddington-limited starburst), indicating that this galaxy must be in a highly transient phase.

Furthermore, the data 
reveal H$_2$O $4_{2,3}-4_{1,4}$ and $3_{3,0}-3_{2,1}$ absorption, with asymmetric blueshifted profiles, which are unambiguous signatures of a massive nuclear outflow, with velocities up to 500~km/s. Using a spherical outflow model, we find a mass outflow rate of $\sim100-900$\,M$_{\odot}$\,year$^{-1}$, implying a mass loading factor $\eta$ much smaller than unity. An outflow mass loading factor of unity is what is observed in most other star forming galaxies and is what is required by models to `regulate' star formation. In this extreme galaxy the outflow, despite being very massive, is not vigorous enough to regulate star formation. As a consequence, star formation will proceed in a runaway mode, at a high rate and high efficiency, consuming all of the available gas in less than $\sim30$\,Myr, and therefore rapidly quenching star formation. If not replenished with fresh gas from the IGM, the galaxy will then evolve into the population of passive galaxies seen already in place at z$\sim$2--4.

Finally, our results show that the previous interpretation of a major merger for the kinematics of this source is untenable. The compact continuum with very dense and warm outflowing gas must be the core of
a single compact starburst galaxy, while emission lines (CO, [CII]) are likely tracing a gaseous disc (or ring) in the host galaxy.

\begin{acknowledgements}
This paper makes use of the following ALMA data: 2013.1.00722.S, 2015.1.00117.S, and 2016.1.01313.S. ALMA is a partnership of ESO (representing its member states), NSF (USA) and NINS (Japan), together with NRC (Canada), MOST and ASIAA (Taiwan), and KASI (Republic of Korea), in cooperation with the Republic of Chile. The Joint ALMA Observatory is operated by ESO, AUI/NRAO and NAOJ. G.C.J. and R.M. acknowledge ERC Advanced Grant 695671 ``QUENCH'' and support by the Science and Technology Facilities Council (STFC). We thank Justin Spilker for useful comments. Calibrated data cubes are available in electronic form at the CDS via anonymous ftp to cdsarc.u-strasbg.fr (130.79.128.5) or via http://cdsweb.u-strasbg.fr/cgi-bin/qcat?J/A+A/
\end{acknowledgements}

\bibliographystyle{aa}
\bibliography{36989corr}

\appendix
\section{Spherical outflow model}
\subsection{Model overview}
In order to convert our observed absorption profile to physical values (e.g. density, velocity, $\dot{M}$), we first consider a spherical outflow with a set inner radius ($\rm R_{IN}$) and thickness ($\rm dR$) around a continuum source of radius $\rm R_{CONT}$. 
Next, we assume that the inner continuum source radiates isotropically, with no limb darkening. This results in three types of observed spectral features: the continuum emission from the inner source, $\rm H_2O$ absorption by shell material between this source and the observer, and $\rm H_2O$ emission from the entire shell. 

We require that the mass outflow rate is constant for all positions in the outflow (i.e. $\rho A v=\rho 4 \pi r^2 v=\mathrm{constant}$; \citealt{scar15,carr18}), which implies $\rho \propto 1/(r^2v)$, and may be stated as:
\begin{equation}
v(r)=v_{o} \left( \frac{r}{R_{IN}}\right)^{\gamma} 
\label{vref}
\end{equation}
\begin{equation}
\rho(r)=\rho_{o} \left( \frac{r}{R_{IN}}\right)^{-\gamma-2} 
\label{mconv}
,\end{equation}
where $\gamma$ is a constant. Adopting Gaussian units, the optical depth of an outflow at radius $r$ for a photon of a given transition of water may be given by (\citealt{sobo60,cast70,proc11,scar15,carr18}):
\begin{equation}
\tau(r)=\frac{\pi e^2f_{lu}\lambda_{lu}\rho_l(r)}{m_ecm_{H_2O}|dv/dr|}=\tau_{o} \left( \frac{r}{R_{IN}} \right)^{-2\gamma-1}
\end{equation}
\begin{equation}\label{tau0}
\tau_o=\frac{\pi e^2f_{lu}\lambda_{lu}\rho_{lo}R_{IN}}{m_ecm_{H_2O}v_o\gamma}
,\end{equation}
where $f_{lu}$ is the oscillator strength, $\lambda_{lu}$ is the wavelength of this transition, and $\rho_l(r)$ is the number density of water molecules in the lower state. We note that this form assumes a large velocity gradient, a central isotropic emitter, and negligible stimulated emission.

The fraction of photons blocked by a parcel of gas is given by:
\begin{equation}
B=1-e^{-\tau(r)}
\label{bref}
.\end{equation}

Each photon from the IR source will only interact with an H$_2$O molecule when the Doppler-shifted frequency of the photon is equal to the resonant frequency of a transition, meaning that the approaching hemisphere of the outflow will be blueshifted, and the receding half redshifted. 
We focus only on the transition of $\rm H_2O\,3_{3,0}-3_{2,1}$.
Because we assume spherical symmetry, the physical space to explore may be described by only two variables: the distance from the centre of the IR source, measured in the plane of the sky ($\rm r_{sky}$), and the position in the shell, as measured parallel to our line of sight ($\rm z_{shell}$).
We split the domain of $\rm r_{sky}$ into small segments, examine each value of $z_{shell}$, and use equation \ref{bref} to calculate the fraction of incident radiation that is absorbed by each cell. We then account for the symmetry of our model by multiplying this cell radiation fraction by the fraction of the continuum source that is subtended by this value of r$_{sky}$ (i.e. $\rm 2 r_{sky} dr/{R_{in}}^2$), and we add this contribution to the total spectrum.

Since the outflow is certainly highly turbulent, we also consider a velocity dispersion $\sigma_v$ by convolving our continuum-normalised spectrum with a 1D Gaussian.

\subsection{Model implementation}

Due to the arguments in the main text, we believe that the rest-frame of the redshift is most likely that traced by the CO-based redshift ($z_{CO}=5.656$), but will also consider the effects of assuming the redshift of the red component of [CII] emission ($z_{[CII],R}=5.6594$; \citealt{litk19}).

While any value of $\gamma$ would satisfy mass conservation (equations \ref{vref} and \ref{mconv}), we
require decreasing velocity and density as a function of radius (i.e. $-2<\gamma<0$), and thus assume $\gamma=-1$.

While it is also possible to fit for the shell thickness ($\rm dR$), these fits yields poor constraints. 
The resolution of the data is insufficient to directly observe the spatial extent of the outflow.
However, detailed modelling of local ULIRGs similar to SPT0346-52 (i.e. very compact) suggest $\rm dR/R_{CONT}$ values of $\sim3$ for NGC4418 and $\sim19$ for Arp220 \citep{gonz17}. 
Since the central cores of both of these sources exhibit the necessary temperature for the upper level of our transitions ($\sim 350-400$\,K), we explore the effects of  the value of $dR/R_{CONT}=3$, $19$, and the intermediate case of $10$.

In order to account for the emission, the four emission profiles of [CII]\,158$\mu$m \citep{litk19}, CO(8-7) \citep{dong19}, CO(2-1) \citep{arav16}, and H$_2$O $2_{0,2}-1_{1,1}$ and $2_{1,1}-2_{0,2}$ (2013.1.00722.S and 2015.1.00117.S, respectively) were averaged and binned to the same velocity resolution as our observations.
This average emission profile was added to the convolved spectrum, with a scaling factor.
However, this scaling factor was highly uncertain, and was usually quite small.
Because of this, we assume that the emission is negligible.

There are several additional parameters that we do not consider, including the ratio of the
continuum source radius to the shell inner radius (i.e. the size of the hollow region inside the shell) and aperture (as a fraction of the total source radius) of our observations. 
As the aperture becomes smaller than unity, the amount of flux that is absorbed at small negative velocities decreases. 
Since we observe a large amount of absorption at these velocities, we  assume that both values are unity.
In addition, it is possible to consider a biconic outflow rather than a spherical outflow, but if azimuthal symmetry is assumed, then any change in the conic opening angle would result in lower absorption at small negative velocities.
In summary, we assume that the inner boundary of the shell is coincident with the outer surface of the continuum source, and consider the entire spherical outflow shell.

We apply this model to the best spectrum for SPT0346-52: H$_2$O $3_{3,0}-3_{2,1}$ for the northern component. 
Using the Bayesian inference code MultiNest \citep{fero08} and its python wrapper (PyMultiNest; \citealt{buch14}), we fit for the following free parameters: the velocity at the interior surface of the shell (v$_{\circ}$), the optical depth constant ($\tau_o$), and the FWHM of the convolving Gaussian ($\sigma_v$).

Since we have no initial information on the geometry of the outflow, our prior probability distributions are relatively broad. We adopt log-uniform distributions for $v_{\circ}$ and $\sigma_v$ ($10^0-10^{3.5}$\,km\,s$^{-1}$), and a log-uniform for $\tau_o$ in the range $10^{-8}-10^0$.

We may estimate the mass outflow rate of the shell by assuming the simple form \citep{rupk05}:
\begin{equation}
\dot{M}= 4\pi r^2 \rho v
.\end{equation}
Due to our assumption of mass conservation in the outflow, this value is constant for all radii inside the outflow. Next, we may estimate the density of water in each lower state by inverting equation \ref{tau0}:
\begin{equation}
\rho_{H_2O,l}(R_{IN})=\frac{\tau_o m_e c m_{H_2O} v_o \gamma}{\pi e^2 f_{lu} \lambda_{lu} R_{IN}}
.\end{equation}
Adopting a density ratio of $X_{H_2O}=\rho_{H_2O}/\rho_{total}$ and assuming a Boltzmann distribution \citep{mang15}, the total density can  be written as:
\begin{equation}
\rho_{total}=\frac{\rho_{H_2O}}{X_{H_2O}}=\frac{\rho_{H_2O,l}Q_{rot}e^{\frac{E_l}{kT}}}{X_{H_2O}g_l}
,\end{equation}
where $\rho_{total}$ is the total matter density of the outflow, $Q_{rot}$ is the rotational partition function,  and $g_l$ is the degeneracy of the lower state. 
We may determine $f_{lu}$ using \citep{hube86}:
\begin{equation}
f_{lu}=(1.4992)\lambda^2A_{ul}\frac{g_u}{g_l}
,\end{equation}
where $\lambda$ is measured in centimetres, $g_u=g_l=7$, and $A_{ul}=0.06616$\,s$^{-1}$.
Combining these expressions, we find:
\begin{equation}
\dot{M}=\frac{4 R_{IN} \tau_o m_e c m_{H_2O}v_o^2Q_{rot}e^{E_l/kT}}{e^2 f_{lu} \lambda_{lu} X_{H_2O}g_l}
.\end{equation}

Since the water absorption is unresolved, we assume an upper size limit of the beam size ($\sim0.13''=0.8$\,kpc). Adopting a magnification factor of $\mu=5.6\pm0.1$ \citep{spil16} for this compact source, this corresponds to a source-plane radius of $R_{IN}=0.07$\,kpc. 

If we assume that the outflow is dominated by H$_2$, then we may approximate $X_{H_2O}\sim\rho_{H_2O}/\rho\sim(m_{H_2O}/m_{H_2})(N_{H_2O}/N_{H_2})\sim9\chi_{H_2O}$, where $\chi_{H_2O}=N_{H_2O}/N_{H_2}$ is the abundance of water with respect to H$_2$. This abundance varies by environment: $\chi_{H_2O}\sim0.5\times10^{-7}$ for star forming regions in the Milky Way \citep{flag13} and $\chi_{H_2O}\sim10^{-6}$ for the warm, extended components of local ULIRGs (\citealt{gonz10,gonz12}), although these values are model dependent. Much higher water abundances of $\chi_{H_2O}\sim10^{-4}$ have been found for shock-heated gas in protostellar outflows \citep{meln08}, but lower abundances have been derived for the outflow as a whole (e.g. $\chi_{H_2O}\sim3\times10^{-7}$; \citealt{kars14,ball16}). Since SPT0346-52 is an extreme ULIRG, we adopt the same abundance inferred for local ULIRGS ($\chi_{H_2O}\sim10^{-6}$).


\end{document}